\documentclass[aps,twocolumn,superscriptaddress,a4paper]{revtex4}
\usepackage{amsfonts,amssymb,amsmath}
\usepackage[utf8]{inputenc}
\usepackage[pdftex,breaklinks=true,colorlinks=true]{hyperref}
\usepackage[pdftex]{graphicx}
\usepackage{epstopdf}
\usepackage{leftidx}
\usepackage{filecontents}

\newcommand{\ket}[1]{\left|#1\right>}
\newcommand{\bra}[1]{\left<#1\right|}

\begin{document}

\title{Dynamics of the spin-1/2 Heisenberg chain initialized in a domain-wall state}

\author{Gr\'egoire Misguich}
\affiliation{Institut de Physique Th\'eorique, Universit\'e Paris Saclay, CEA, CNRS, F-91191 Gif-sur-Yvette, France}
\author{Kirone Mallick}
\affiliation{Institut de Physique Th\'eorique, Universit\'e Paris Saclay, CEA, CNRS, F-91191 Gif-sur-Yvette, France}
\author{P. L. Krapivsky}
\affiliation{Institut de Physique Th\'eorique, Universit\'e Paris Saclay, CEA, CNRS, F-91191 Gif-sur-Yvette, France}
\affiliation{Department of Physics, Boston University, Boston MA 02215, USA}

\date{\today}

\begin{abstract}
We study the dynamics of an isotropic spin-1/2 Heisenberg chain starting in a domain-wall initial condition, where the spins are initially up on the left half-line and down on the right half-line. We focus on the long time behavior of the magnetization profile. We perform extensive time-dependent density-matrix renormalization group simulations (up to $t=350$) and find that the data are compatible with a diffusive behavior.
Subleading corrections decay slowly blurring the emergence of the diffusive behavior. 
We also compare our results with two alternative scenarios:  superdiffusive behavior and enhanced  diffusion with a logarithmic  correction. We finally discuss the evolution of the entanglement entropy.
\end{abstract}

\maketitle

\section{Introduction}

Many-body systems far from equilibrium led to many fascinating theoretical ideas and experimental breakthroughs.  Despite the remarkable advances during the last decades \cite{Polkovnikov2011,Derrida2007,Bertini2014}, a fundamental framework for non-equilibrium statistical physics is still under intense development.  Some progress has been achieved by studying  toy models amenable  to numerical or analytical analyses, such as classical interacting particle systems,  one-dimensional quantum spin chains and cold quantum  gases \cite{SpohnBook,TakahashiBook,SutherlandBook}. Luckily, these mathematical models  turn out to be relevant for diverse experimental fields ranging from cold atoms and magnetism to soft-condensed matter and biophysical transport \cite{Bloch2004,Kinoshita2006,Takeuchi2011,Dalibard2012,Chou2011}.

In classical physics,  a  thorough understanding of elementary systems  such as exclusion processes or solid-on-solid  growth models \cite{Derrida2007,HalpinHealy1995,Sasamoto2008} has provided us with key insights on
far from equilibrium fluctuations  \cite{Jona-Lasinio2014}, large deviations \cite{TOUCHETTE}, persistent influence of the
initial conditions \cite{Gerschenfeld2009,CorwinRV} and hydrodynamic limits \cite{Derrida2011}.

For  quantum many-body systems, far from equilibrium  thermalization, integrability,   initial preparation of the system  are 
fundamental riddles  on which the
recently proposed~\cite{castro-alvaredo_emergent_2016,bertini_transport_2016} generalized hydrodynamics may shed new light (see, e.g., \cite{collura_analytic_2017,Vasseur2017,Vasseur2017b,doyon_large-scale_2017,ilievski_microscopic_2017,piroli_transport_2017,doyon_geometric_2018} and references therein). At its present stage, the generalized hydrodynamics approach has been successfully applied to integrable one-dimensional systems characterized by ballistic transport. It is usually impossible to find explicit solutions of infinitely many coupled Bethe-Boltzmann equations underlying the generalized hydrodynamics, but very precise results have been obtained e.g. using iteration procedures \cite{castro-alvaredo_emergent_2016,bertini_transport_2016,collura_analytic_2017,Vasseur2017,Vasseur2017b,doyon_large-scale_2017,ilievski_microscopic_2017,piroli_transport_2017,doyon_geometric_2018}. 

Dynamics of integrable systems with sub-ballistic transport remain more challenging even on the conceptual level. The isotropic Heisenberg spin chain is the first integrable many-body quantum system which is in principle solvable by the Bethe ansatz~\cite{TakahashiBook,SutherlandBook,Bethe1931}, but its transport properties are still beyond the reach of exact calculations. Here we investigate the isotropic Heisenberg spin chain initialized in a domain wall initial condition. Spin transport in this system has been extensively studied and it has been claimed that the magnetization profile displays an anomalous superdiffusive scaling behavior (see~\cite{ljubotina_spin_2017,ljubotina_class_2017} and references therein).  The aim of the present work is to revisit this issue using large-scale numerical simulations. To our knowledge our data are among the most precise and extensive available at the moment, and they support the simplest interpretation, namely the diffusive transport. 
The diffusive behavior is difficult to extract  due to large subleading corrections. The same phenomenon was also noticed in a very recent calculation of the return probability after a quench in the  Heisenberg spin chain~\cite{stephan_return_2017}.

\section{The model}

We study  the evolution of an XXZ spin-1/2  chain, 
  initialized at time $t=0$ in a domain wall configuration
$\ket{ \uparrow\uparrow\cdots\uparrow\uparrow\downarrow\downarrow\cdots\downarrow\downarrow}$ where all the spins in the left half of the system are  ``up'' ($S^z=\frac{1}{2}$) and those in the right half are ``down'' ($S^z=-\frac{1}{2}$).
At time $t>0$ the wave function of the system is then defined
by
\begin{equation}
 \ket{\psi(t)}=\exp\left(-iHt\right)\ket{ \uparrow\uparrow\cdots\uparrow\uparrow\downarrow\downarrow\cdots\downarrow\downarrow},
\end{equation}
where  the Heisenberg Hamiltonian $H$ is that of a XXZ chain of length $L$ with open boundary conditions,
\begin{equation}
  H=\sum_{r=-L/2}^{L/2-2} \left(S^x_r S^x_{r+1}+S^y_r S^y_{r+1} +\Delta S^z_r S^z_{r+1} \right), 
\end{equation}
and $\Delta$ is the anisotropy parameter.
We  will focus in particular on the long time behavior of the magnetization profile: $m(r,t)=\bra{\psi(t)} S^z_r \ket{\psi(t)}$~\footnote{It is easy to check that the sign of $\Delta$, as well as that in front of the $xy$ terms, are irrelevant for this problem \cite{gobert_real-time_2005}, and we therefore take  $\Delta\geq0$.}.

This problem was first studied by Antal {\it et al.}~\cite{antal_transport_1999,antal_logarithmic_2008} in the free fermion case ($\Delta=0$), where an exact analytical solution for the long-time limit of the magnetization profile was  obtained. A few years later the problem with $\Delta\ne0$ was  studied numerically by Gobert {\it et al.} \cite{gobert_real-time_2005} using the time-dependent density-matrix renormalization group (DMRG) (see also~\cite{sabetta_nonequilibrium_2013}). For $\Delta<1$, the  numerical results   implied that the 
magnetization satisfies  the following scaling at long times:
$m(r,t)=f(r/t)$. This  ballistic propagation of the magnetization front for $\Delta<1$  is by now well established,
and  has been  confirmed by the calculation,  in  the long time limit,
of  the magnetization profile $f$, using  the
  hydrodynamic equations derived from the thermodynamic  Bethe Ansatz  for  the XXZ model~\cite{castro-alvaredo_emergent_2016,bertini_transport_2016,collura_analytic_2017}.
The velocity of the front is given by the simple formula   $v_f = \sqrt{ 1 - \Delta^2}$ \cite{collura_analytic_2017,stephan_return_2017}. The velocity 
vanishes when $\Delta \to 1^-$, and, when  $\Delta>1$, the behavior is completely different, and the magnetization profile freezes at long time \cite{gobert_real-time_2005,MosselCaux2010}.
This should not be confused with the situation where the system is initialized in
a mixed state where both halves of the chains are {\em partially} polarized ($\langle S^z\rangle_{\rm left/right}=\pm\mu/2$ with $\mu<1$).
With such an initial state, the dynamics is diffusive for $\Delta>1$ \cite{ljubotina_spin_2017}. 

The situation in the isotropic case ($\Delta=1$) is much  less clear but  particularly interesting.
The  early results of Gobert {\it et al.} \cite{gobert_real-time_2005} indicated that the spreading of the magnetization profile  obeys  a power law $\sim t^\alpha$, with an exponent $\alpha$ around $0.6$  (since this exponent is greater than $\frac{1}{2}$, the system is said to be  superdiffusive). More precisely, the authors of   \cite{gobert_real-time_2005} observed  that the magnetization profiles at long times can be described by the simple scaling form  $m(r,t)\simeq g(r/t^\alpha)$, and that the total magnetization (or charge) $Q(t)=\sum_{r=0}^{L/2} \left[m(r,t)+\frac{1}{2}\right]$ transferred from the left  to the right since $t=0$ was increasing  proportionally to  $t^\alpha$.
Similar results have been  obtained by Ljubotina {\it et al.}~\cite{ljubotina_spin_2017,ljubotina_class_2017}, who,  using simulations up to $t=200$, predict  an exponent $\alpha$  in the range between $0.6$ and $\frac{2}{3}$. These authors  argue that a superdiffusive behavior is also substantiated  by transport properties of the anisotropic Heisenberg model \cite{Affleck2011}.
 
On the other hand,  St\'ephan has recently computed \cite{stephan_return_2017} the return probability $\mathcal{R}(t)=\left|\left< \psi(t) | \psi(0)\right>\right|^2$ and showed that
\begin{equation}
\mathcal{R}(t)\sim \sqrt{t}\exp\left(-\gamma \sqrt{t}\right), \;\;{\rm where}\;\;  \gamma=\zeta(3/2)/\sqrt{\pi}.
\label{eq:Return}
\end{equation}
As argued in~\cite{stephan_return_2017}, this result~\footnote{Note that the very same formula has also appeared for a the symmetric exclusion process in Ref.~\cite{kms2015}.} is 
incompatible  with the  exponent $\alpha$   larger than $\frac{1}{2}$: If the front spreads as $t^\alpha$, 
 then  the overlap of $\ket{\psi(t)}$ with the initial domain wall is expected to satisfy   $\mathcal{R}(t) 
 \lesssim  {\rm e}^{-at^\alpha}$. This implies $\alpha \le 1/2$.
We also note that Collura {\it et al.}~\cite{collura_analytic_2017} analyzed the magnetization profile in the vicinity of the edge of the light cone
when $\Delta\to 1^-$, and this led them to conjecture a diffusive behavior at  $\Delta=1$.

We have  performed numerical simulations of the Heisenberg spin-1/2  chain up to 
 the time  $t=350$. 
We will show that, assuming a pure power law (as in Refs.~\cite{gobert_real-time_2005,ljubotina_spin_2017,ljubotina_class_2017}), our data up to $t=350$ indicate that $\alpha$ is  smaller than $0.6$. More interestingly, the same data are perfectly  compatible with a diffusion-like exponent $\alpha=\frac{1}{2}$, provided one includes some subleading corrections  that vanish in the long-time limit.  Finally, we also probe the validity of the third scenario, where the diffusive behavior  is marginally enhanced by multiplicative logarithmic correction term.

\section{Numerical results}

\begin{figure}[h]
\includegraphics[height=0.48\textwidth, angle=270]{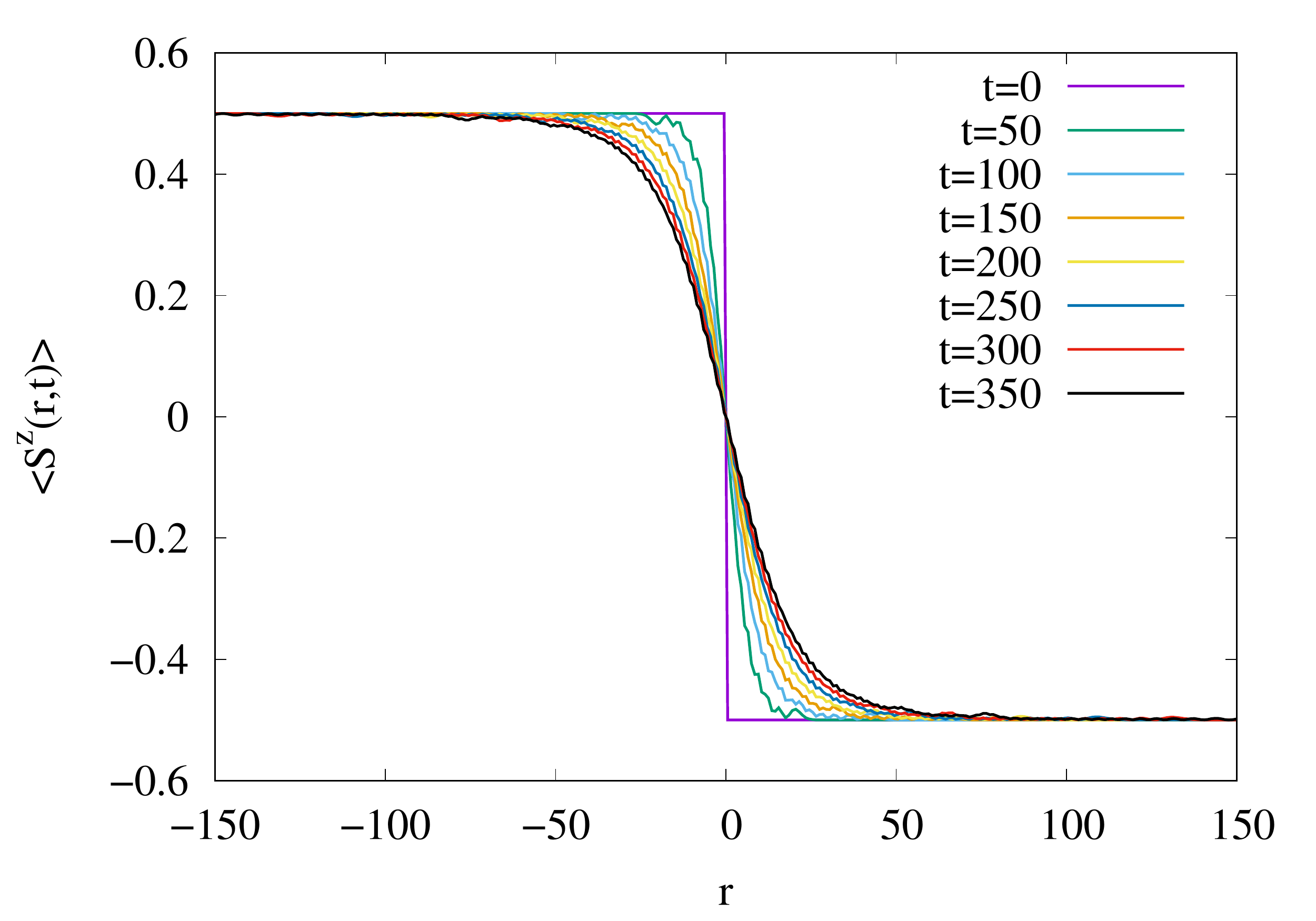}
\caption{Magnetization profiles at different times. Simulation parameters: maximum bond dimension $\chi=2000$, Trotter step $\tau=0.3$, and
system size $L=800$ sites (only 300 sites in the center are shown   here).
}
\label{fig:profile}
\end{figure}

\begin{figure}[h]
\includegraphics[height=0.48\textwidth, angle=270]{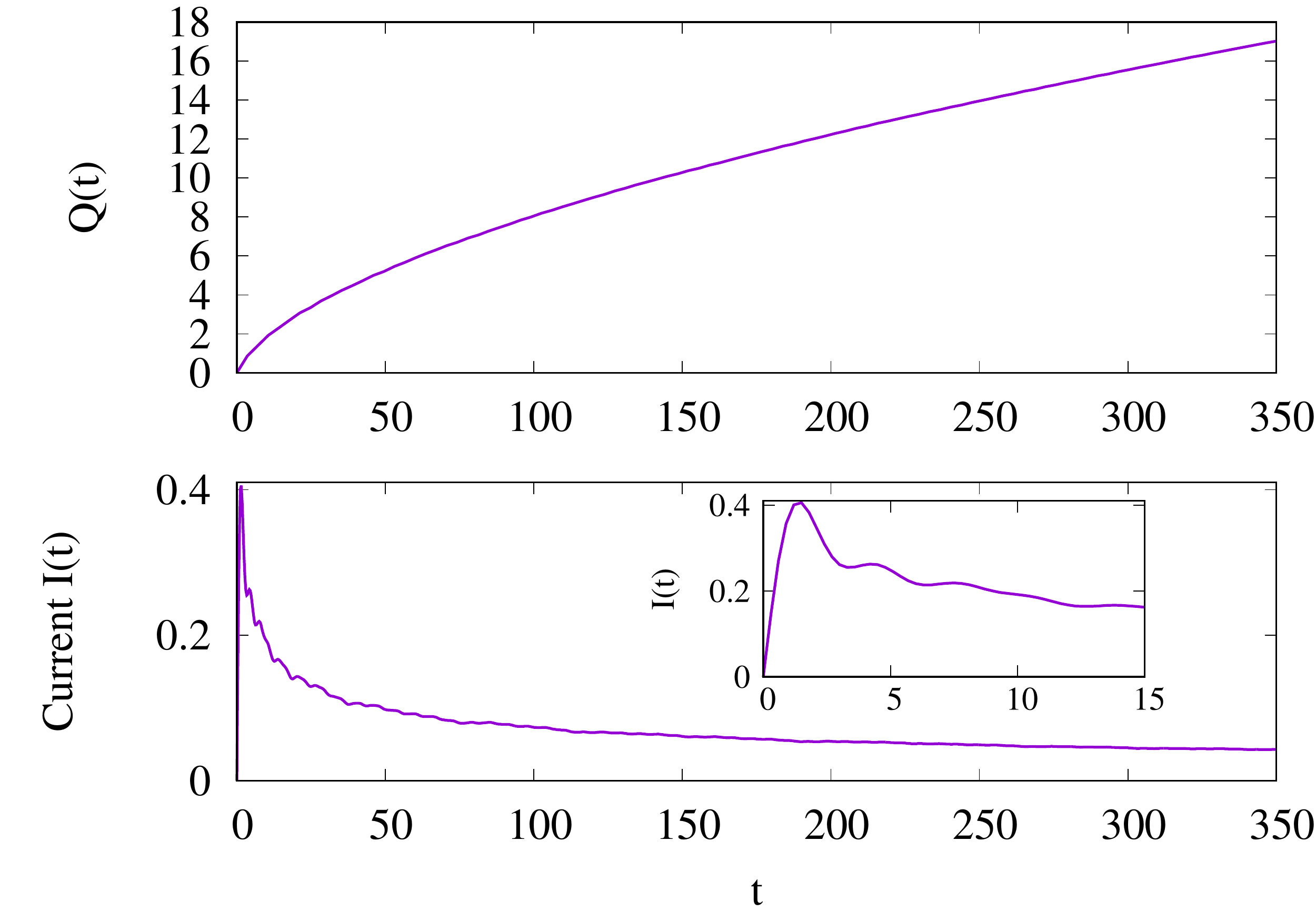}
\caption{
Top: Magnetization $Q(t)$ transferred from the left to the right since $t=0$.
Bottom: current $I(t) = \frac{ d}{dt}Q(t)$.
Simulation parameters: maximum bond dimension $\chi=2000$, Trotter step $\tau=0.3$, and
system size $L=800$ sites.
}
\label{fig:DI}
\end{figure}

In  Fig.~\ref{fig:profile}, we  show  the evolution of the magnetization profile up to $t=350$, a final time significantly larger than  the  time $t=200$  reached in the simulations  of Ref.~\cite{ljubotina_class_2017}. 
In order to analyze quantitatively how this profile spreads in time,
we plot the magnetization $Q(t)$ transferred from the left to the right since $t=0$ (top of Fig.~\ref{fig:DI}), and  its time derivative, the current 
$I(t)=i\bra{\psi(t)} S^+_0S^-_1 - S^-_0S^+_1 \ket{\psi(t)}$  measured in the center
of the chain (bottom of Fig.~\ref{fig:DI}).  Some details about the numerical
  method are given in Appendix~\ref{app:dmrg}.

\subsection{Extracting an effective exponent: Superdiffusive behavior?}

\begin{figure}[h]
\includegraphics[height=0.48\textwidth, angle=270]{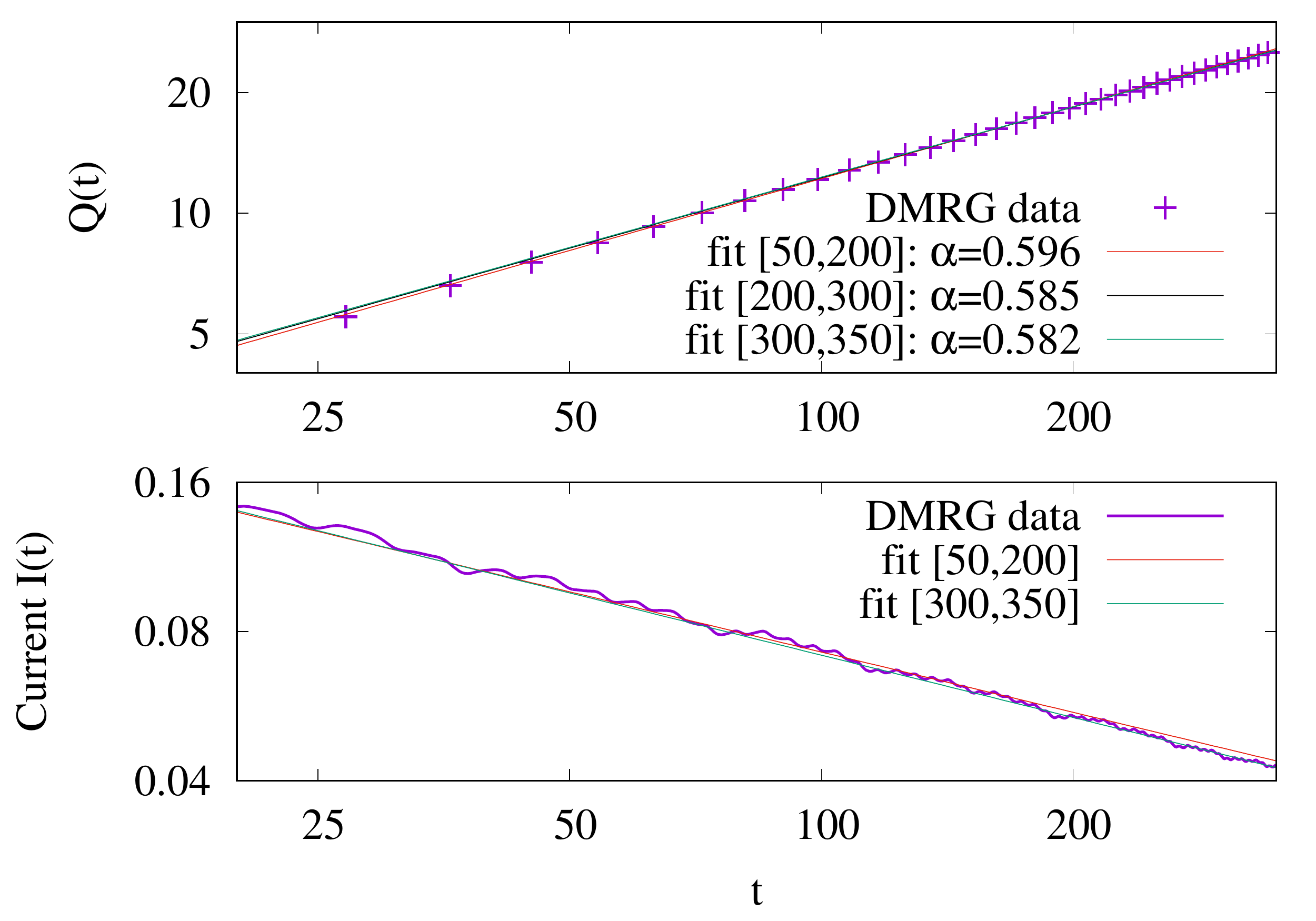}
\caption{
Same data as in Fig.~\ref{fig:DI}, displayed in a log-log scale.
Top: the data are fitted  to some power law $Q(t)\sim t^\alpha$.
Depending on the time window used for the fit ($[50,200]$, $[200,300]$, or $[300,350]$,
we get $\alpha$ between $\simeq 0.596$ and $0.582$ (see also Tab.~\ref{tab:alpha}). Bottom: current $I(t)$ is compared with the derivative of the power laws obtained in the above fits.
}
\label{fig:DIlog}
\end{figure}

We start by performing an analysis similar to that
of Refs.~\cite{gobert_real-time_2005,ljubotina_spin_2017}, where $Q(t)$ is fitted using a simple power law:
$Q(t)\simeq t^\alpha$. The results of three different fits are shown in Fig.~\ref{fig:DIlog},
where the data are plotted using a log-log scale.
Depending on the time window used for the fit ($[50,350]$, $[200,350]$, or $[300,350]$),
we extract  a value of  $\alpha$ between $\simeq 0.596$ and $0.58$ (more details in Tab.~\ref{tab:alpha}). We emphasize that  this effective  exponent  decreases with times, and  its true value is therefore very likely  to  be smaller than the  value  $\frac{3}{5}$ proposed in 
\cite{ljubotina_spin_2017,ljubotina_class_2017}. It is  plausible that the effective value of $\alpha$
 would decrease if one could perform simulations that last even longer (going over $t=400$ is
 difficult for the moment). Besides, the current exhibits oscillations that hinder a precise
 determination of the exponent. These facts encouraged us to look for alternative interpretations
 of our numerical data and to examine whether  it could be compatible with a more orthodox diffusive
 behaviour.

\subsection{Diffusion with  subleading corrections}

We now show that the data obtained from the DMRG simulations  are perfectly  compatible with a diffusive exponent $\alpha=\frac{1}{2}$, provided one includes some correction terms in the long-time expansion.  We shall  discuss two possibilities: (i) a  subleading $1/t$ correction in $I(t)$, or (ii) some multiplicative logarithm.

\begin{figure}[h]
\includegraphics[height=0.48\textwidth, angle=270]{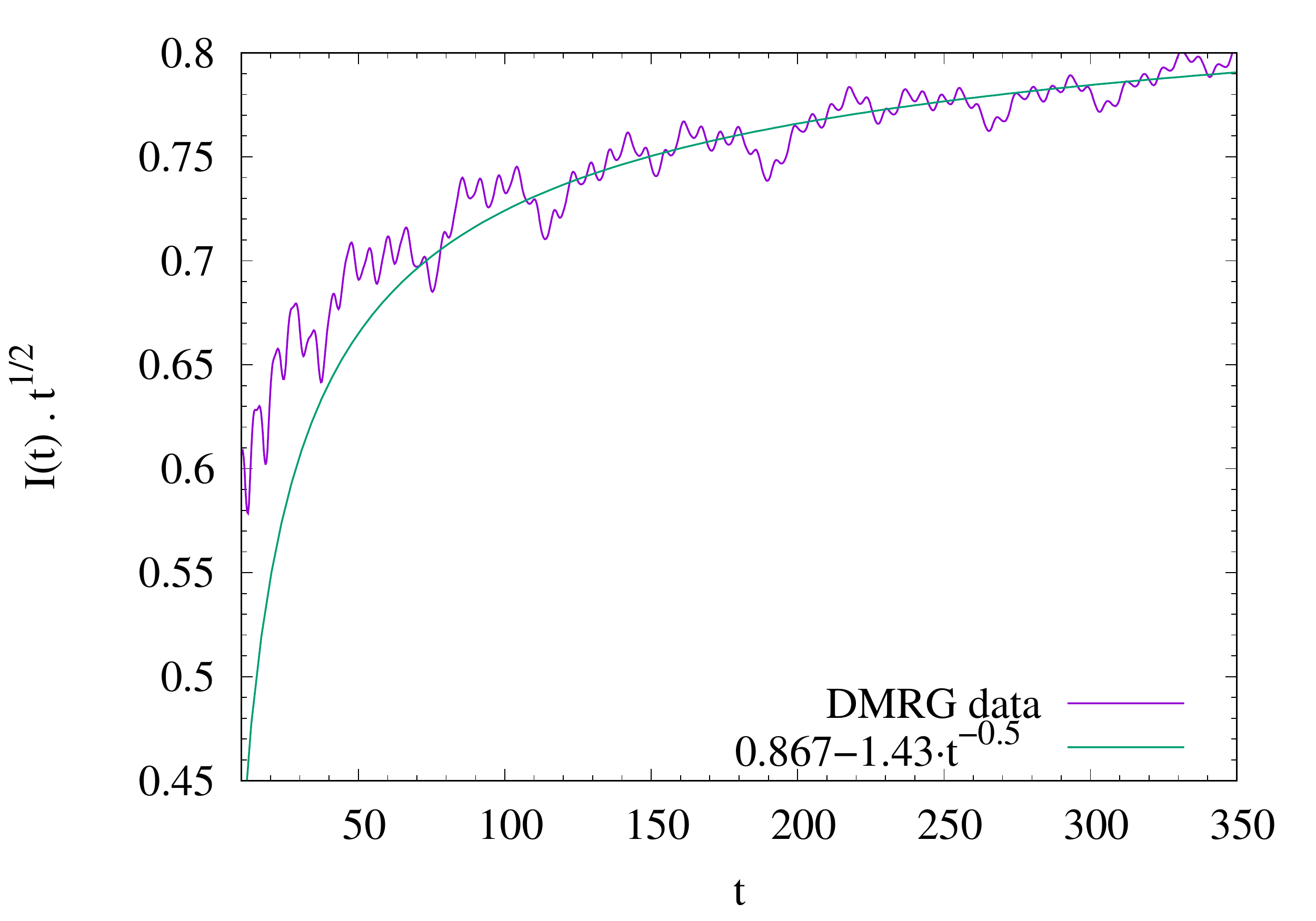}
\caption{Current $I(t)$ multiplied by $\sqrt{t}$ to highlight the long-time part.
The data appear to be well fitted by a function of the type $a-bt^{-1/2}$ (green line).
Fit window: $[200:350]$.
}
\label{fig:I_0.5}
\end{figure}

The  current $I(t)$  is not given by  a pure power law for any finite range  of $t$. First, as is clear from  Fig.~\ref{fig:I_0.5},
there is an oscillatory behavior (for a more quantitative discussion of these  oscillations see  Appendix~\ref{app:osc}).
Besides, even if the oscillations are averaged out, there remain subleading corrections to the dominant asymptotic contribution to  $I(t)$.
Conjecturing  a normal diffusive scenario,  it is  natural to  expect that $I(t)$ will admit some long-time expansion in  powers
of $t^{-\frac{1}{2}}$ of the form $I(t) =t^{-\frac{1}{2}}(a + {b}t^{-\frac{1}{2}} +  c t^{-1} + \ldots).$
Keeping only the first two terms in this expansion, we observe in  Fig.~\ref{fig:I_0.5}  that the current $I(t)$ is very well  approximated by the form
\begin{equation}
I(t)\simeq at^{-\frac{1}{2}}-bt^{-1},
\end{equation}
 with $a \approx 0.867$ and $b \approx 1.43$.  Equivalently, by integration, we obtain
 that the total magnetization transferred exhibits a logarithmic correction to the dominant
 $\sqrt{t}$ behavior, 
\begin{eqnarray}
Q(t)\simeq R(t), \quad R(t)=2a t^{\frac{1}{2}}-b\ln(t) 
\label{eq:R}
\end{eqnarray}

\begin{figure}[h]
\includegraphics[height=0.48\textwidth, angle=270]{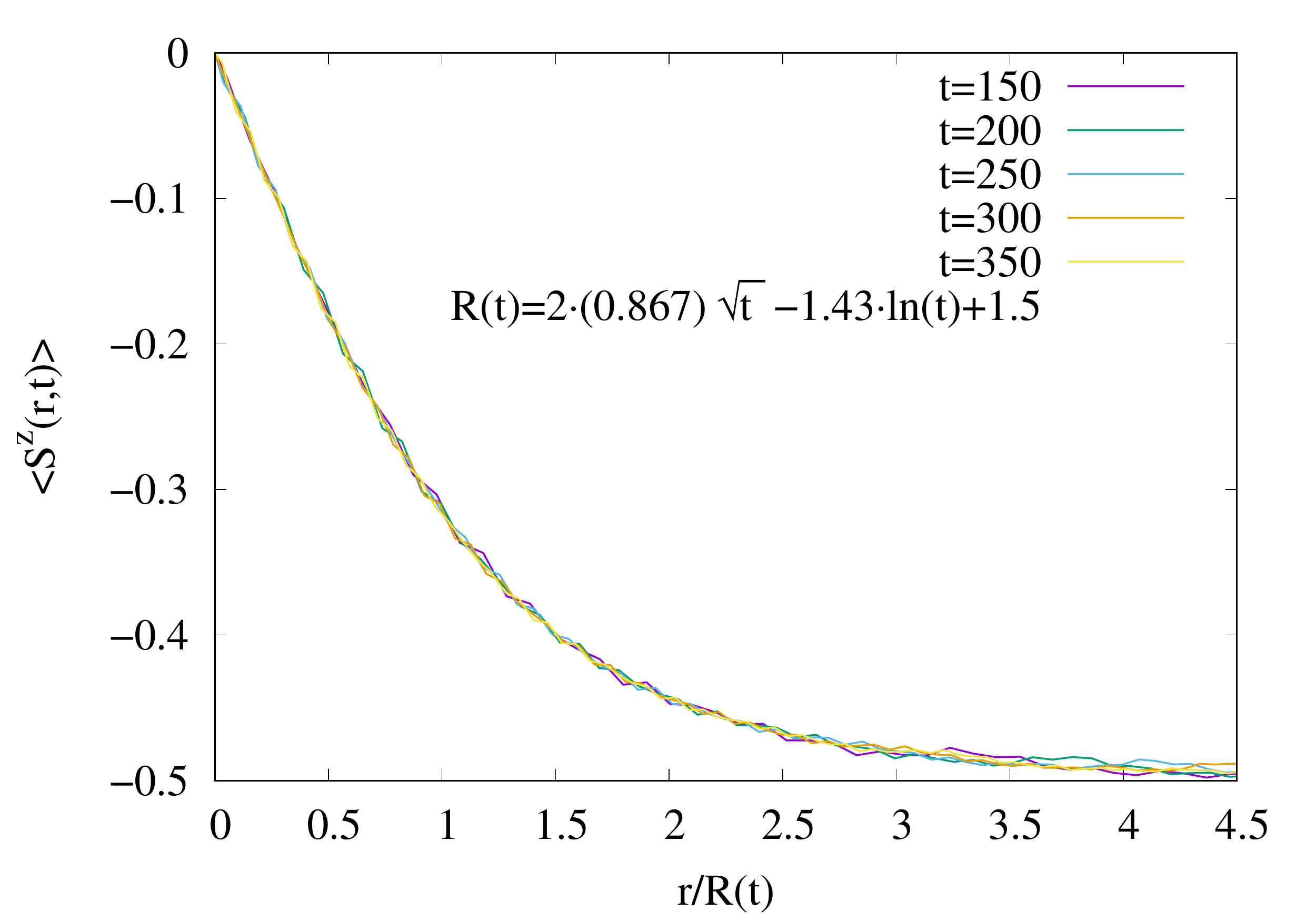}
\caption{Magnetization profiles at different times, plotted as a function of the rescaled distance $r/R(t)$, where  $R(t)$ is  defined in Eq.~\eqref{eq:R}.
The coefficients $a$ and $b$ used to define $R(t)$ are those obtained by fitting the current $I(t)$ in Fig.~\ref{fig:I_0.5}.}
\label{fig:profile_sqrt}
\end{figure}

It is interesting to compare this result with the return probability
given by Eq.~\eqref{eq:Return}.  As explained in \cite{stephan_return_2017}, the quantity  $l(t) = -\ln[\mathcal{R}(t)]$ can be seen as a typical {\it length scale}  over which the initial state and $\ket{\psi(t)}$ differ. This gives  $l(t)\sim \gamma \sqrt{t} -\frac{1}{2}\ln(t)+\mathcal{O}(1)$, which also  includes a subleading logarithmic term, as is the case  for $R(t)$ in Eq.~\eqref{eq:R}. Figure \ref{fig:profile_sqrt} indeed shows that a  good collapse of the magnetization profiles is obtained if the  distance $R(t)$ is used as a dynamical length scale. 

\begin{figure}[h]
\includegraphics[height=0.48\textwidth, angle=270]{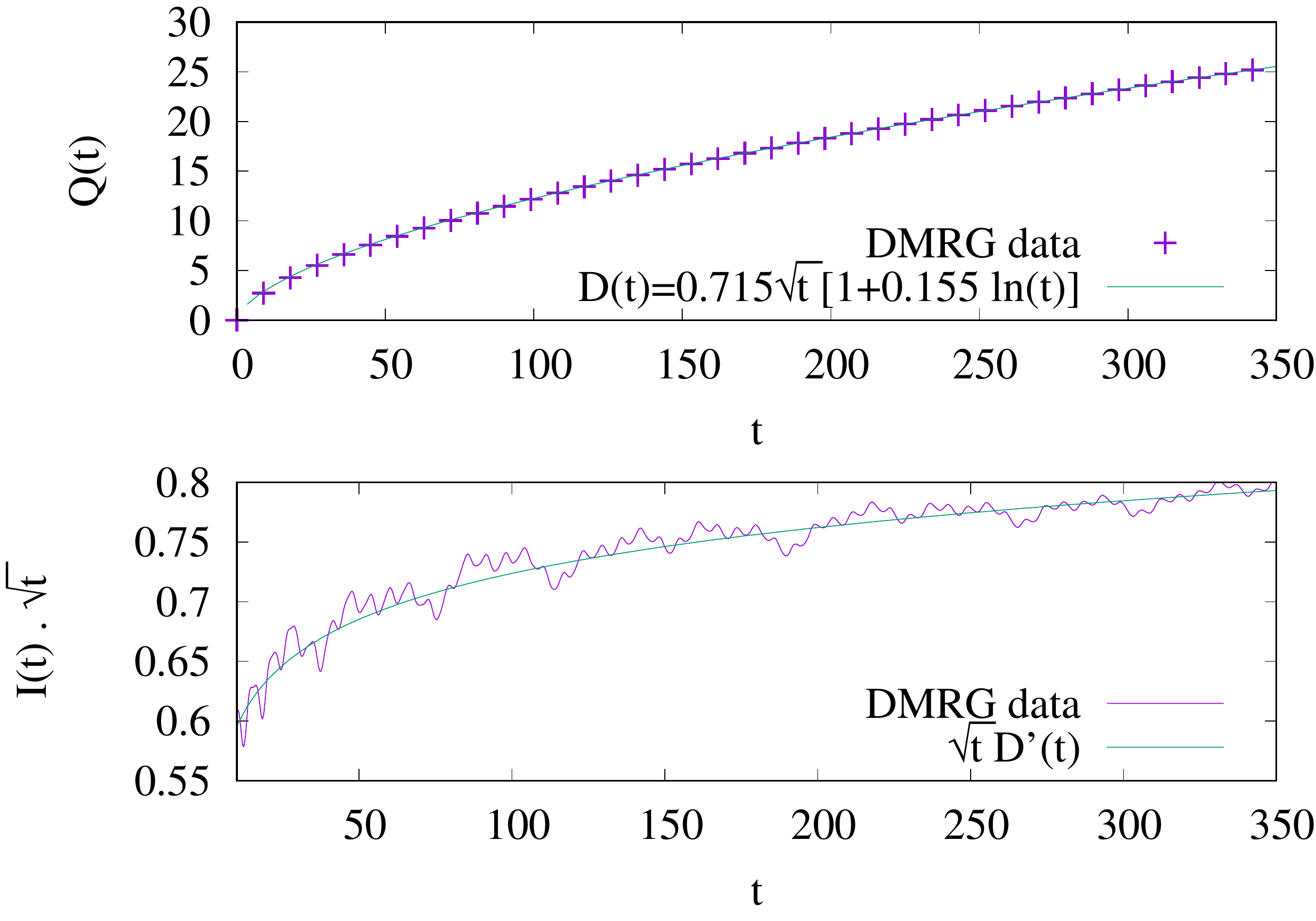}
\caption{Top: transferred magnetization $Q(t)$ and fit to a diffusive behavior corrected by a multiplicative logarithm $D(t)=c  \sqrt{t} \left[1+d \ln(t)\right]$
(fit window $[200:350]$).
Bottom: Current $I(t)$ multiplied by $\sqrt{t}$ and comparison with the derivative of the function $D(t)$ above.
}
\label{fig:I_0.5log}
\end{figure}

Finally, we  consider a last  scenario, where the transferred magnetization has $\sqrt{t}$ behavior that is corrected by a multiplicative  logarithmic factor:
\begin{equation}
Q(t)\simeq c \sqrt{t} \left[1+d \ln(t)\right],
\label{eq:Dlog}
\end{equation}
which is equivalent to
\begin{equation}
I(t)\simeq \frac{1}{2} c  t^{-\frac{1}{2}} \left[1+2d + d \ln(t)\right]. 
\label{eq:Iaveclog}
\end{equation}
In that case, the profile does  display a superdiffusive behaviour, but with a marginal enhancement.
The result of a fit using Eq.~\eqref{eq:Dlog} is shown in Fig.~\ref{fig:I_0.5log}, and this expression 
also  matches  the data quite well.  Having extracted empirical values for  $c$ and $d$ from the fit [Eq.~\eqref{eq:Dlog}], we checked  the accuracy  of  Eq.~\eqref{eq:Iaveclog}, see  Fig.~\ref{fig:I_0.5log} (bottom): the agreement is fairly good.

\subsection{Entanglement entropy}
\label{ssec:S}

\begin{figure}[h]
\includegraphics[height=0.48\textwidth, angle=270]{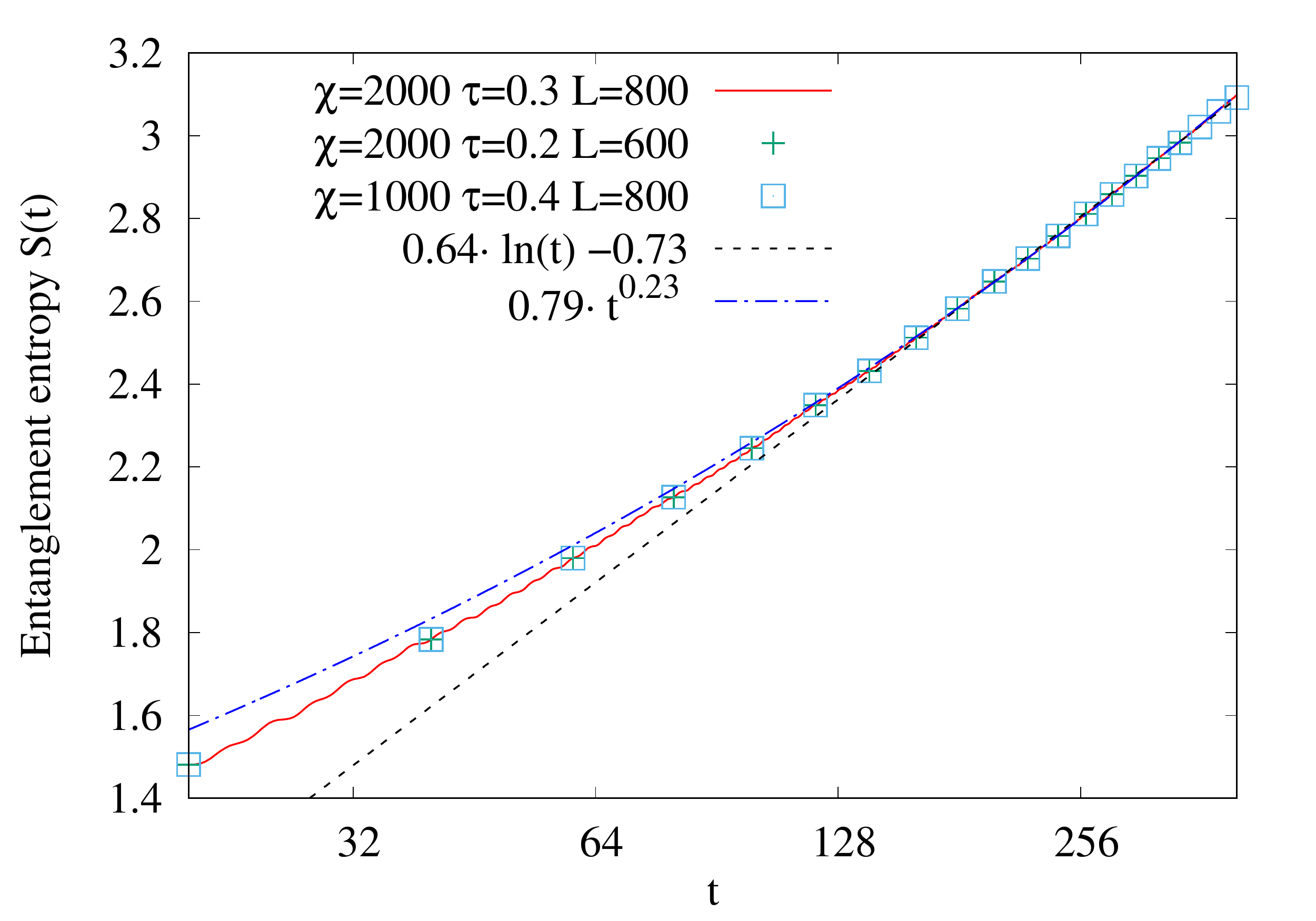}
\caption{Entanglement entropy $S(t)$ as a function of time, for a bipartition in the center of the chain.
The results for three different sets of simulation parameters  are displayed
($\chi$: maximum bond dimension, $\tau$: Trotter step and $L$: system size). The black dashed line
is a fit to $\sim \ln(t)$. The blue one is a fit to a power law, giving an exponent $0.23$
which is relatively close to that ($0.25$) proposed in Ref.~\cite{ljubotina_spin_2017}.
Both fits are performed using the data for $t\geq 150$.
}
\label{fig:S}
\end{figure}

The entanglement entropy of a subsystem $A$ is defined by $S(t)=-{\rm Tr_A}\left[\rho_A(t) \ln \rho_A(t)\right]$,
where $\rho_A(t)={\rm Tr}_B \left[|\psi(t)\rangle\langle\psi(t)|\right]$
is the reduced density matrix of the region $A$, and is obtained by tracing out the spins in the complement $B$ of the region $A$.
Figure~\ref{fig:S} shows $S(t)$, the entanglement entropy of the left half of the chain. 
The results obtained with different simulation parameters are compared, and the small differences between them turn out to be practically invisible at the scale of the figure.
This indicates that the chosen parameters (maximum bond dimension $\chi$, Trotter time-step $\tau$ and system size $L$) provide a good precision up to the largest times reached in these calculations. 

Concerning the long time behavior of $S(t)$, we observe (Fig.~\ref{fig:S}) that the data are compatible with a logarithmic growth as well as with a power law.
A logarithmic entropy growth is a common behavior after a local quench in a one-dimensional critical system, and this can be understood using conformal field theory methods~\cite{calabrese_entanglement_2007,stephan_local_2011,eisler_entanglement_2012}. It was however argued in Ref.~\cite{ljubotina_spin_2017} that, in the present case,
the entropy grows algebraically,  $S\sim t^\beta$ with $\beta \approx 0.25$.
We indeed find that a power law with a  small exponent seems to  reproduce the data over 
a larger time window than $S \sim \ln(t)$, but it nevertheless is difficult to draw a  firm conclusion from the available numerical data.
Investigating the full counting statistics (and its relations to entanglement~\cite{klich_quantum_2009,song_bipartite_2012,song_entanglement_2011,eisler_full_2013}), might be a way to make progress on this question.

\begin{figure}[h]
\includegraphics[height=0.48\textwidth, angle=270]{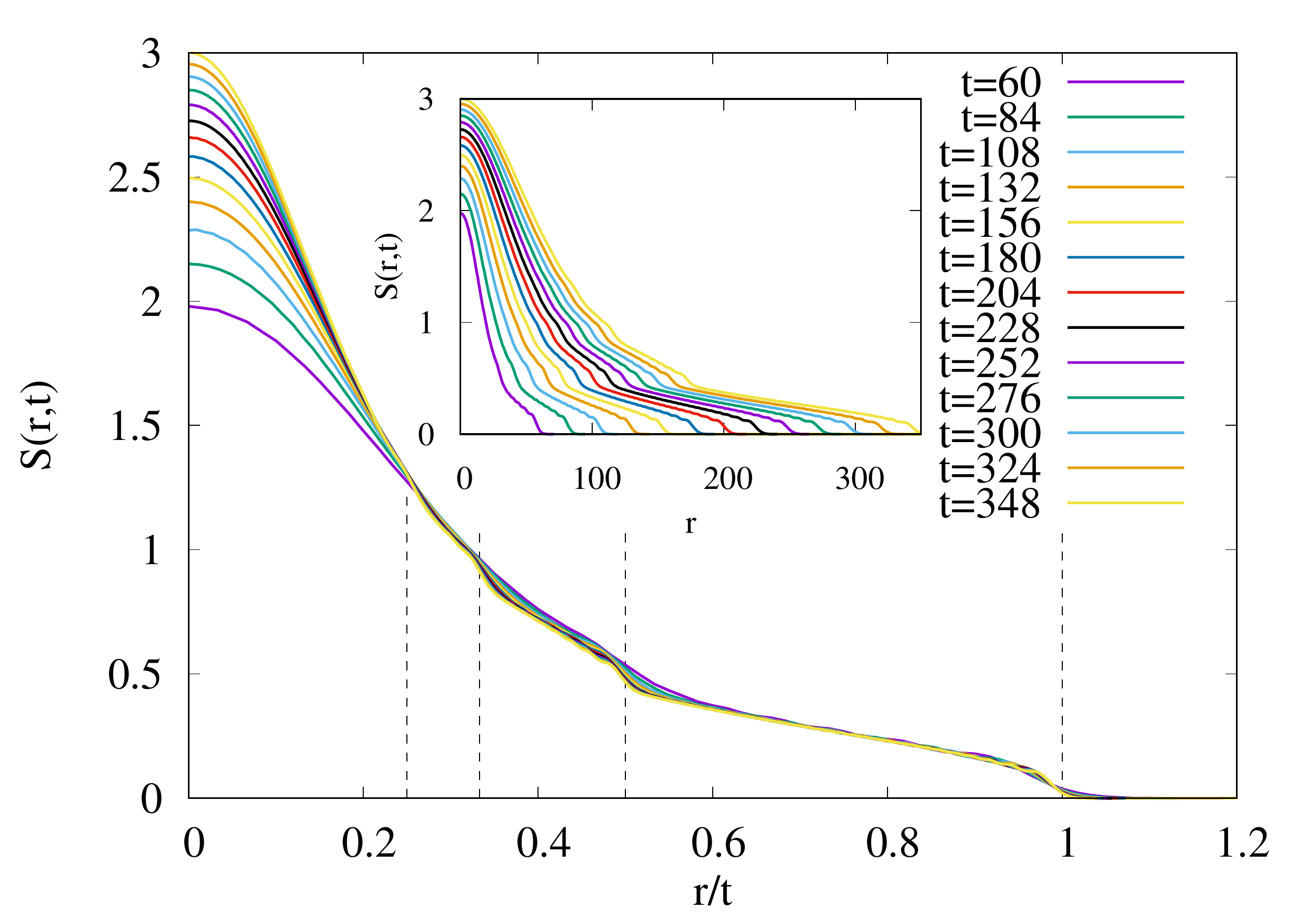}
\caption{Entanglement entropy $S(t,r)$ as a function of time $t$ and position $r$ of the cut.
Inset: same data plotted as a function of the `bare' position $r$.
Simulation parameters: maximum bond dimension $\chi=2000$, time-step $\tau=0.3$, and system size $L=800$. The dashed vertical lines,
located at $r/t=1,1/2,1/3$ and $1/4$,
are guides to the eye.}
\label{fig:Sprof}
\end{figure}

We also consider $S(t,r)$, the entanglement entropy associated with a  left-right partition of the chain performed at position $r$ (the origin $r=0$ being here the center of the bond located in the middle of the chain).
The resulting entropy profiles are displayed in Fig.~\ref{fig:Sprof}. A first observation is that, using the rescaled position $r/t$, the data
obtained at different times approximately collapse on a single curve, at least for $r/t\gtrsim 0.3$. So,
contrary to the magnetization
which spreads relatively slowly in time, and certainly not in a ballistic way, the entanglement entropy
is well fitted by $S(t,r)\simeq s(r/t)$ at sufficiently long times.
We also note that the tip of the entropy profile, at $r/t\simeq 1$, corresponds to the maximum
group velocity $v=1$ of a single magnon in a ferromagnetic background, as also noted in \cite{stephan_return_2017}.
It is also intriguing to note some shoulder-like structures around $r/t\simeq 0.5$, $r/t\simeq 0.33$ and possibly around $0.25$ too.
These could be related to the propagation of some magnon bound states, as discussed in Refs.~\onlinecite{ganahl_observation_2012,vlijm_quasi-soliton_2015} in the context of a different quench in the XXZ spin chain. 
Closer to the center of the chain, the entropy profiles at different times  clearly do not overlap.
For the largest times shown in Fig.~\ref{fig:Sprof}, this happens for $r/t\lesssim 0.3$.
It turns out that this corresponds to the spatial region of width $\sim \sqrt{t}$ where the magnetization deviates significantly from $\pm \frac{1}{2}$.
So, we may expect that the entropy data should collapse
for $r/t\gg t^{-\frac{1}{2}}$.  
Since the entropy $S(t,r=0)$
in the center diverges with time (Fig.~\ref{fig:S}), the existence of a limiting profile $s(x)$ with a divergence at $x=0$ seems to be a plausible scenario.

\section{Discussion}

We have analyzed the magnetization profile and the spin current in the isotropic Heisenberg spin-1/2 chain starting from a quench where the system is initially prepared in a domain-wall product state.

Although many quantities can be computed at thermal equilibrium, no analytical calculation of the shape of the evolving magnetization profile in the infinite system is known. Even the scaling with time of the typical size of the profile is unknown and subject to
controversy. Whereas in the anisotropic case, the profile is either ballistic (for $\Delta <1$) or
frozen (for $\Delta > 1$), it is not clear whether the isotropic point $\Delta = 1$  displays normal
diffusion  or is superdiffusive with respect to time.
 
Recent numerical simulations have been interpreted in favor of superdiffusive behavior with an exponent close to 3/5.   We have performed large-scale  numerical
simulations  indicating that the effective exponent, evaluated over a finite window of time, is smaller
than this value. Moreover, we show that the numerical  data  can be very well interpreted in favor of  normal
diffusion  behavior provided that subleading corrections to the dominant behavior (which are known
always to exist) are taken into account. This  interpretation  implies   that the dynamical length scale
grows as the square-root of the time with a logarithmic correction,  in agreement with the exact calculation
of the return probability. Although it may be possible to improve the numerical simulations to  reach even 
larger times, we believe that time is ripe for  analytical investigations of this vexing problem, either
by using integrability or by studying some effective and simplified  models and comparing predictions
with the  accurate  numerical data that the DMRG method and its variants allow us to gather. 

\section*{Acknowledgements}
We are grateful to Vincent Pasquier and Jean-Marie Stéphan for several useful discussions. K.M. thanks S.~Mallick
 for a careful reading of the manuscript. 

\appendix
\section{Details about the DMRG simulations}
\label{app:dmrg}

Our calculations are performed using a time-dependent DMRG algorithm, implemented using the C++ iTensor library~\cite{itensor}.
The evolution operator $U=\exp(-i\tau H)$ for a time-step $\tau$ is approximated by a matrix-product operator~\cite{zaletel_time-evolving_2015}, using a 4-th order Trotter scheme.
Unless specified otherwise, we use a time-step $\tau=0.3$, system size  $L=800$ and a matrix-product state representations of $\ket{\psi(t)}$ with matrices of
size up to $\chi=2000$~\footnote{In the regions where the entropy is sufficiently low, the dimension of each matrix can be fixed by demanding that the discarded weight in the singular value decomposition is smaller than $10^{-10}$. When such a dimension becomes larger than $\chi$, we limit the matrix size to $\chi$.}.

\begin{figure}[h]
\includegraphics[height=0.48\textwidth, angle=270]{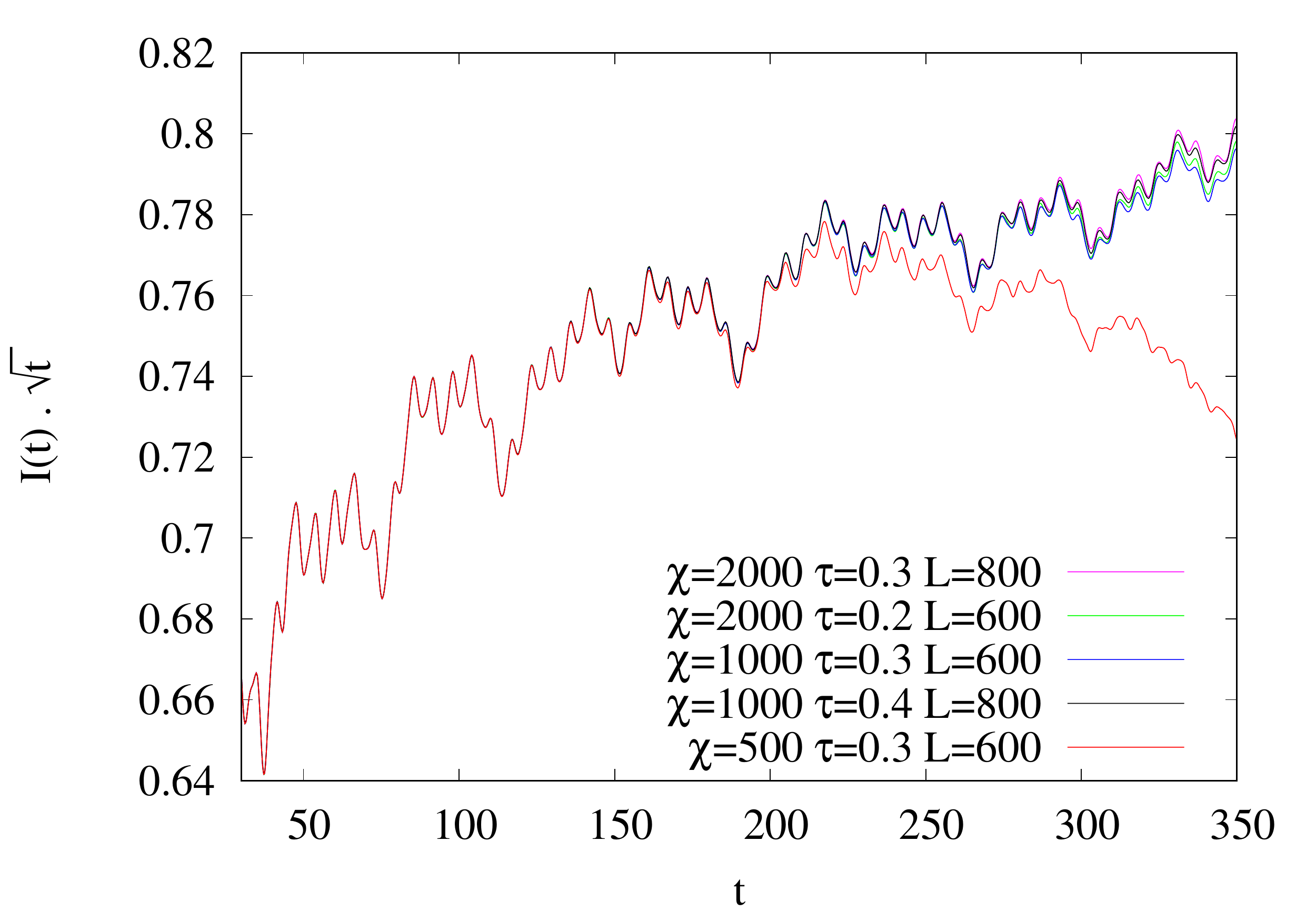}
\includegraphics[height=0.48\textwidth, angle=270]{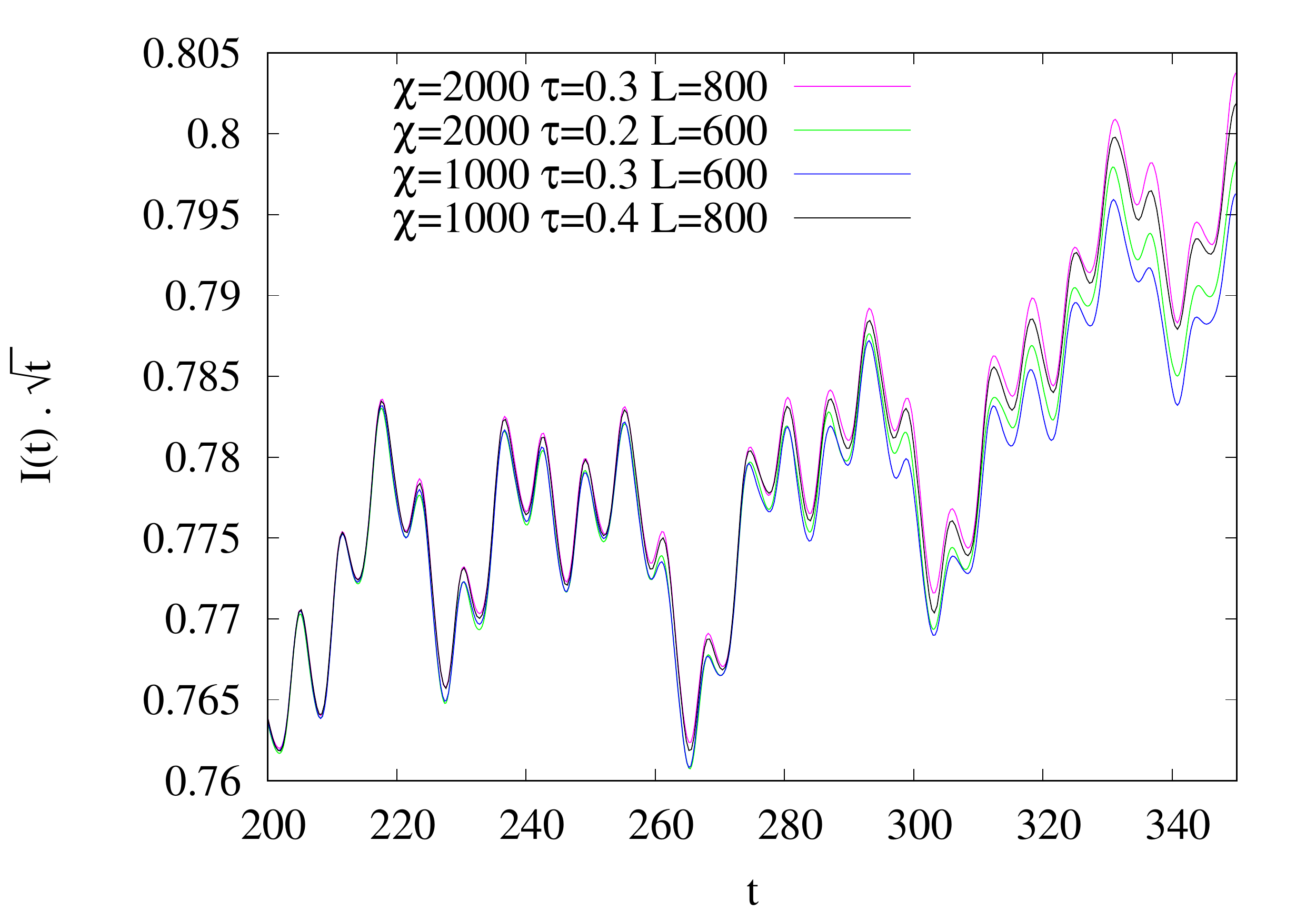}
\caption{
Current $I(t)$ multiplied by $\sqrt{t}$.
The precision of the DMRG simulation is checked by varying some parameters: maximum bond dimension $\chi$, Trotter
step $\tau$ and system size $L$. The bottom panel is a zoom on the large-time part of the data.
Note that the value $\chi=500$ (with $\tau=0.3$, red curve in the top panel) appears to be too small to achieve a sufficient precision beyond $t\sim 200$.
On the other hand, the calculations with $(\tau,\chi)=(0.3, 1000)$, $(0.4, 1000)$, $(0.2,2000)$ and $(0.3,2000)$ match up to $t\simeq 200$. They give very close results up to $t\simeq 300$ (green, blue and black curves differ by less
than 0.5\% in relative value), and stay relatively close to each other up to $t=350$ (relative differences below $1\%$).
}
\label{fig:check_presicion}
\end{figure}

To check the accuracy,  calculations with some smaller $\chi$, as well as with 
 various   values of $\tau$ and $L$ were performed.
As can be seen in Fig.~\ref{fig:check_presicion}, the different simulations agree very well at the scale of the figures  up to $t\simeq 300$, and their relative differences stay smaller than $1\%$ up to the longest time, $t=350$.
The precision of the results can also be judged by
looking at how the estimated value of the exponent $\alpha$
depends on the simulation parameters $\tau$, $\chi$ and $L$.
As shown in Tab.~\ref{tab:alpha}, these effects are relatively small.
Finally we note that the entanglement entropy is often quite sensitive to truncations errors in DMRG simulations, and
the good convergence observed in Fig.~\ref{fig:S}  for different simulations parameters
is also a good check of the precision of the results.

\begin{table}
\begin{tabular}{|c|c|c|c|c|c|}
\hline
$\alpha$ & fit error & time window & $N$  & $\tau$ & bond dim. $\chi$ \\
\hline
0.5980 & $\pm 0.00030$		& $[50,150]$	& 800	& 0.3	& 2000 \\
0.5958 & $\pm 0.00023$ 		& $[50,200]$ 	& 800 	& 0.3 	& 2000 \\
0.5958 & $\pm 0.00051$		& $[50,200]$ 	& 800 	& 0.2 	& 3000 \\
0.5851 & $\pm 0.00024$ 		& $[200,300]$ 	& 800 	& 0.3 	& 2000 \\
0.5845 & $\pm 0.00036$		& $[200,300]$ 	& 600	& 0.3	& 1000	\\
0.5825 & $\pm 0.00040$		& $[300,350]$	& 800	& 0.3	& 2000 \\
0.5819 & $\pm 0.00037$		& $[300,350]$	& 800	& 0.4 	& 1000 \\
0.5805 & $\pm 0.00056$		& $[300,350]$	& 600	& 0.2 	& 2000 \\
0.5798 & $\pm 0.00053$		& $[300,350]$	& 600	& 0.3	& 1000	\\
\hline\end{tabular}
 \caption{Variations of the fitted exponent $\alpha$ (obtained from $Q(t)$ in a log-log scale, as in Fig.~\ref{fig:DIlog}) with respect to the DMRG simulation parameters. The second column is  the standard error from the least-square fit (does not take into account the possible variations
 with $\chi$ and $\tau$). It should be recalled that decreasing the Trotter step $\tau$ at fixed $\chi$ does not necessarily gives more precise calculations, as it implies more frequent matrix truncations along the time evolution.}
 \label{tab:alpha}
\end{table}

\section{Current oscillations}
\label{app:osc}
\begin{figure}[h]
\includegraphics[height=0.48\textwidth, angle=270]{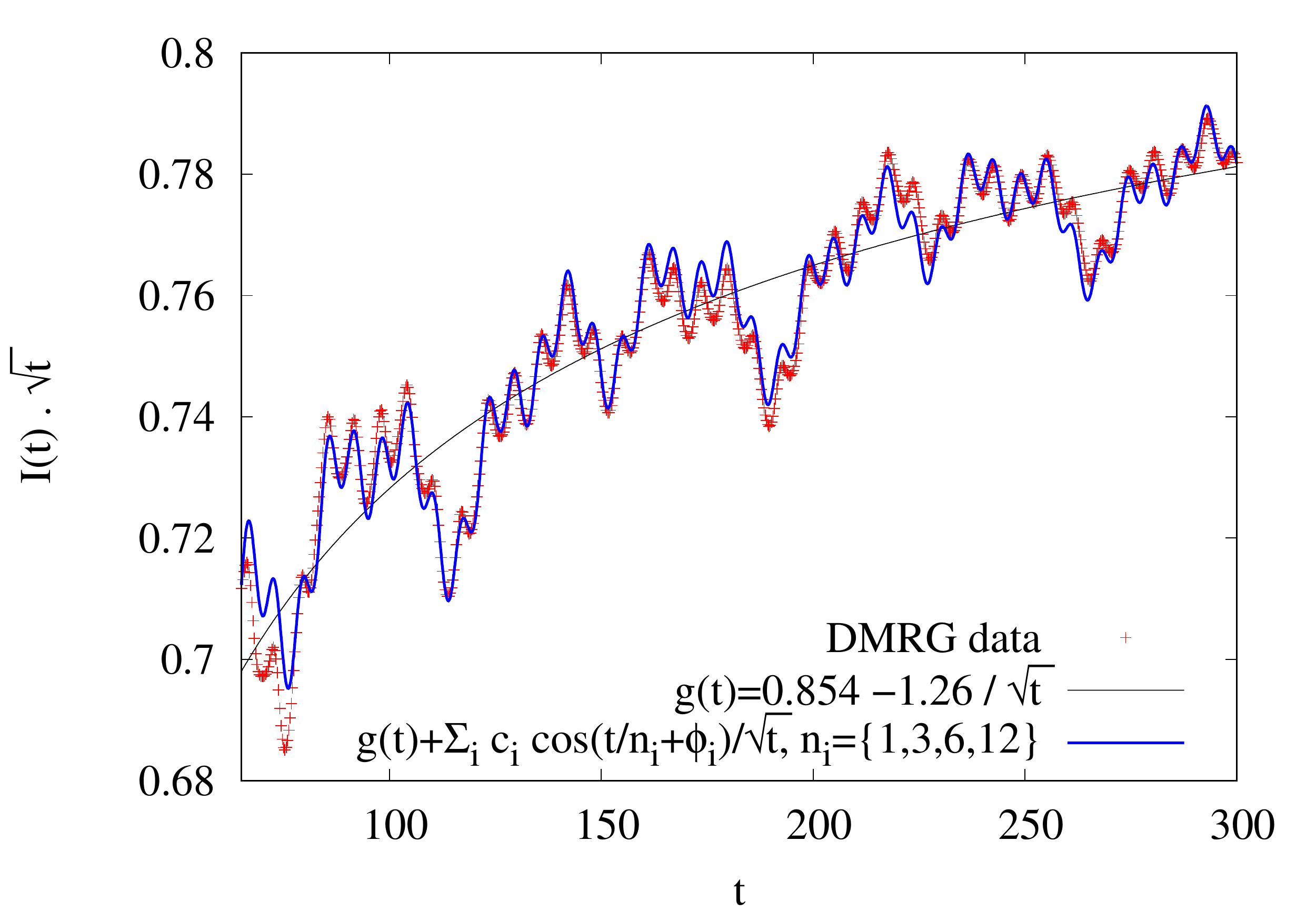}
\caption{Current $I(t)$ multiplied by $\sqrt{t}$ (red crosses) and fitting function given in Eq.~\eqref{eq:osc} (blue line). The black
line represents the terms in the fitting function which do not oscillate.
}
\label{fig:osc}
\end{figure}

As already noticed in Ref.~\cite{stephan_return_2017}, the current $I(t)$ displays some oscillations. We observe that their amplitudes decay slowly with time, in a way which is compatible with a $1/t$ behavior.
In addition, the signal is dominated by a few harmonics with periods  $\Delta t=2\pi$, $6\pi$, $12\pi$ and $24\pi$.
To make these remarks concrete, we have fitted  the data using the following function
\begin{eqnarray}
 I(t) &\simeq& \frac{a}{\sqrt{t}} + \frac{1}{t}\left[b + c_1\cos(t+\phi_1) \right. \nonumber \\ 
 &&+ c_3\cos(t/3+\phi_3)+ c_6\cos(t/6+\phi_6)\nonumber \\
 &&\left.+c_{12}\cos(t/12+\phi_{12})\right],\label{eq:osc}
\end{eqnarray}
and the result  is displayed in  Fig.~\ref{fig:osc}.
We note that dropping the $c_{12}$ term provides a relatively good fit too (data not shown), while including an additional $\cos(t/24)$ term makes it even better.
As expected, the numerical values we obtain for $a$ and $b$ are close to those  obtained without the oscillatory terms, in Fig.~\ref{fig:I_0.5}.
It should finally be noted that the (shortest) period $\Delta t=2\pi$ of the first cosine term is a natural time in the problem since it corresponds to the energy change $\Delta E=1$ induced by one spin flip in the ferromagnetic state: \mbox{$|\cdots\uparrow\uparrow\textcolor{red}{\uparrow}\uparrow\uparrow\cdots\rangle\to|\cdots\uparrow\uparrow\textcolor{red}{\downarrow}\uparrow\uparrow\cdots\rangle$}. 

\bibliography{\jobname.bib}{}
\end{document}